\documentstyle{l-aa}

\def\gsimeq
{\hbox{\raise0.5ex\hbox{$>\lower1.06ex\hbox{$\kern-1.07em{\sim}$}$}}}

\begin{document}

   \thesaurus {11.05.2;  
               11.14.1;  
               11.17.3;  
               12.04.2;  
               13.07.2   
               } %
   \title{Can Flat Spectrum Radio Quasars make most of the 
          overall $\gamma$--ray background ?}

   \author{A. Comastri\inst{1}, 
           T. Di Girolamo\inst{2} and
           G. Setti\inst{2,3}
          }

   \offprints{A. Comastri}

   \institute{$^1$ Osservatorio Astronomico di Bologna, via Zamboni 33
              I-40126 Bologna, Italy \\
              $^2$ Istituto di Radioastronomia del CNR, via Gobetti 101, I-40129
               Bologna, Italy\\
              $^3$ Dipartimento di Astronomia, Universit\`a di Bologna,
               via Zamboni 33 I-40126 Bologna, Italy\\
             }

   \date{Received 27 October 1995; accepted March 1996}

   \maketitle

   \markboth{A.Comastri et al. \, FSRQ and the $\gamma$--ray background}
            {A.Comastri et al. \, FSRQ and the $\gamma$--ray background}

   \begin{abstract}

The contribution of discrete sources to the $\gamma$--ray background
is modeled. Flat spectrum radio quasars are known to make a substantial
 contribution
to the hard (E $>$ 100 MeV) background.
The so called ``MeV bump", however,
cannot be accounted for in terms of known classes of sources even
taking into account the newly discovered class of objects with a
a broad band spectrum sharply peaked in the MeV range (``MeV blazars").
Even in the most optimistic case the predicted intensity falls short of at
least a factor three.

   \end{abstract}

%

\section{Introduction}

The most important discovery of the CGRO EGRET in the field of the
extragalactic astronomy is the detection of high energy $\gamma$--rays
(E $>$ 100 MeV) from active galaxies.
At present (Thompson et al. 1995) some 50 sources have been identified.
All these objects appear to emit most of their bolometric luminosity
in $\gamma$--rays and are strong, core-dominated, flat spectrum
($\alpha \leq 0.5$ at a few GHz; $F_{\nu} \propto \nu^{-\alpha}$)
extragalactic radio sources.
The majority have been identified with quasars
while about ten are classified as BL Lac objects.
\par
Even if the precise value of the isotropic $\gamma$--ray background
(GRB)
above a few tens of MeV is still uncertain because of the
need to subtract the galactic contribution,
there is now strong evidence
that the flat spectrum radio quasars (FSRQ)
can supply most of the extragalactic background radiation
above 30 MeV, with smaller fractional contributions
from  BL Lacs and normal and starburst galaxies
(Setti \& Woltjer 1994; Erlykin \& Wolfendale 1995).
\par
The origin of the ``MeV bump" remains a major puzzle. Contrary
to earlier suggestions the recent results on Seyfert spectra (Johnson et al.
 1994)
indicate that they cannot provide any major contribution
to the MeV band. On the other hand it is very difficult to
conceive a physical mechanism taking place at cosmological
distances which can account for the large amount of energy involved
in the ``MeV bump" and, at the same time, being
consistent with the very sharp drop of the background spectrum
from several MeV down to 30 MeV.
\par
Therefore, we have investigated the possibility that the ``MeV
bump" could be accounted for in terms of the summed contribution
from FSRQ, or a fraction thereof, which are known to possess hard
X--ray spectra.
While a straight extrapolation of the X--ray spectra
($<\alpha> \simeq 0.5$) to higher energies cannot provide
any major contribution to the ``MeV bump", the recent discovery by
COMPTEL of several FSRQ (usually referred to as ``MeV blazars") with enhanced
 emission in the MeV band,
well above the extrapolation of their X--ray spectra, may provide
some hope that a subclass of FSRQ could account for the
``MeV bump" (Bloemen et al. 1995).
\par
The purpose of this work is to estimate the contribution from FSRQ
 to the GRB using the available
$\gamma$-ray properties recently discovered by CGRO COMPTEL and EGRET
observations coupled with the available informations
at radio and X-ray wavelengths. Throughout this paper the values
$H_0$ = 50 km s$^{-1}$ Mpc$^{-1}$ and $q_0$ = 0 have been used.

%
%

%

\section{The model}

Our model for the synthesis of the overall GRB is anchored to the
X--ray emission properties of FSRQ at 1 keV. The parameters for the broad
band X-- and $\gamma$--ray spectrum and cosmological evolution of FSRQ
have been chosen in order to obtain a good fit to the overall set of
observational constraints. All the assumed parameters are consistent,
within the errors, with those suggested by the present available
observations.
\par
The local emissivity at 1 keV of $9.7\times 10^{19}$ W Hz$^{-1}$
Gpc$^{-3}$ can be derived from the emissivity at 5 GHz 
of $5.6\times 10^{26}$ W Hz$^{-1}$ Gpc$^{-3}$, which is 
consistent within 1
$\sigma$ with that obtained from FSRQ luminosity 
function by Maraschi \& Rovetti (1994),
by applying an average broad band spectral index 
$<\alpha_{rx}>$ = 0.88.
The adopted value of $\alpha_{rx}$
has been derived from the 1 keV fluxes of a large sample
of FSRQ observed with 
the {\it Einstein} IPC (Wilkes et al. 1994) and in the ROSAT All
Sky Survey (Brinkmann et al. 1994, 1995), and with the 5 GHz fluxes 
reported by Stickel et al. (1994).
[Here and in the following the broad band spectral indices 
$\alpha_{12}$ are defined as $- log(L_2/L_1)/log(\nu_2/\nu_1)$
where $L_1$ and $L_2$ are the rest--frame luminosities observed at 
the frequencies $\nu_1$ and $\nu_2$].

\par
For a typical FSRQ a broken power law spectrum with
$\alpha_{x}$ = 0.5 and $\alpha_{\gamma}$ = 1.2 has been assumed.
The adopted $\alpha_{x}$ is an extrapolation from the mean $\alpha_{x}$
value of the FSRQ 
detected by the {\it Einstein}
IPC (Wilkes et al. 1994, Wilkes \& Elvis 1987) and at higher energies 
by EXOSAT (Comastri et al. 1992), while 
$\alpha_{\gamma}$ is 
the mean value in the EGRET band (Thompson et al. 1995). The break
energy is defined by the requirement that the local emissivity at 100 MeV
is such that, with the cosmological evolution parameters described below,
one can account for a fixed fraction of the high energy GRB.

\par
For the FSRQ pertaining to the ``MeV blazars'' type we have adopted a
broken power law spectrum with $\alpha_{x} = 0$,
$\alpha_{\gamma} = 2$ and a break energy at 2.5 MeV.
These spectral parameters are in good agreement with the recent
COMPTEL observations of a few of these peculiar objects
(Williams et al. 1995, Bloemen et al. 1995).
\par
The evolution is parameterised as a power law such that 
the luminosity $L(z) = L(0) \times (1+z)^{\beta}$. The determination of
evolutionary properties of the $\gamma$--ray sources is made
difficult by their small number and variability. Chiang et al.
(1995) have found, using the $V/V_{max}$ test,
a best fit evolutionary parameter $\beta = 3.0^{+0.5}_{-0.7}$
assuming $q_0$=0.1. This value is
consistent with those found for FSRQ in other regions of the
elecromagnetic spectrum: $\beta \simeq 3$ in the radio band 
(Dunlop \& Peacock 1990) and $\beta \simeq 2.5 - 3$ in the X--ray band 
(Della Ceca et al. 1994). 
We have therefore adopted $\beta = 3$. Following the results
of Dunlop \& Peacock (1990), Boyle et al. (1993) and Warren et al. (1994) the
evolution is cut--off at a redshift $z_{cut}=2.5$, then the emissivity is
assumed to be constant up to $z_{max} = 5$, and zero for larger redshifts.

\par
The predicted background intensity in the energy range 1 keV -- 1 GeV
has been computed assuming that 95\% of the local emissivity at 1 keV
is due to the FSRQ and the remaining 5\% to the ``MeV blazars''.
This ratio reflects the fact that only a few ``MeV blazars" have been
discovered from COMPTEL compared with the roughly 40 FSRQ in the
EGRET band.

   \begin{figure*}
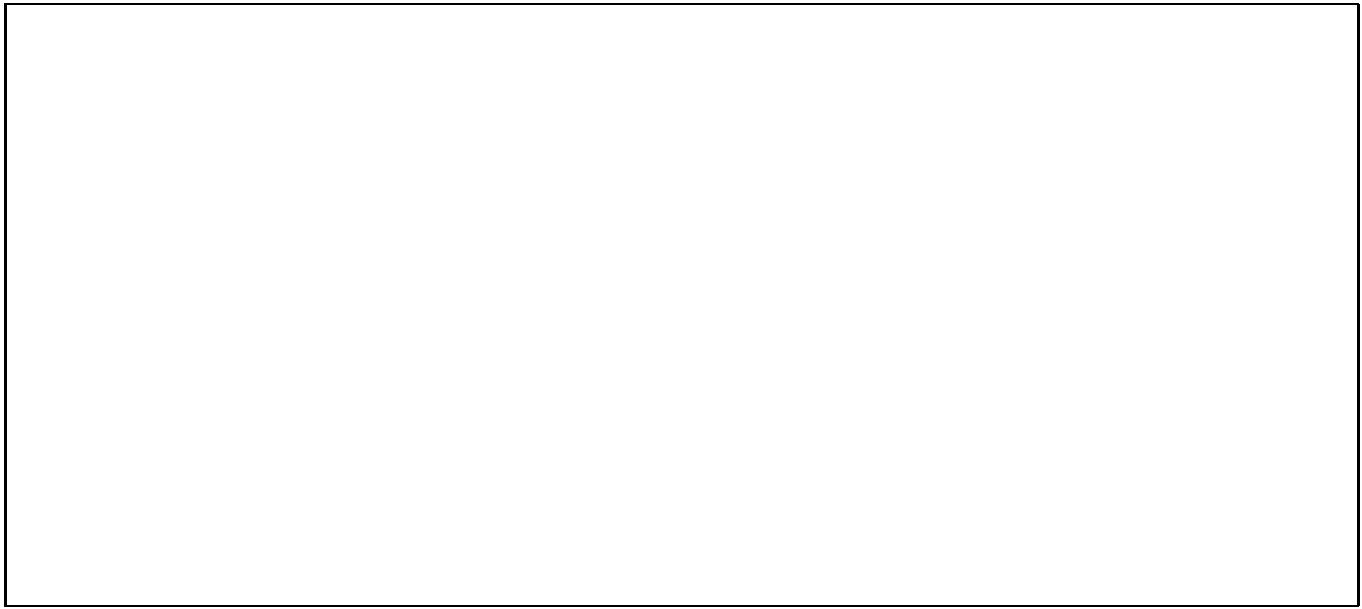

     \picplace{8cm}
      \caption{The predicted X-- and $\gamma$--ray
               backgrounds (solid line) with the three distinct contributions 
               (dashed lines) of FSRQ, ``MeV blazars" and of the AGN
               model of Comastri et al. (1995; labeled as ``XRB fit").
               The thick error bars in the MeV region represents
               the recent results obtained by COMPTEL (see Kappadath et al. 
               this conference). The high energy GRB points above 30 MeV
               are from Osborne et al. (1994).
              }

         \label{FigGam}%
    \end{figure*}

\par
In Figure 1 we have reproduced a selection of data on the X-- and
 $\gamma$--ray backgrounds. 
For the high energy GRB we have plotted the results from the 
SAS--2 data analysis of Thompson \& Fichtel (1982) and those obtained by
Osborne et al. (1994) on the basis of the EGRET Phase 1 data from the
CGRO archives. A more recent analysis of the EGRET data has been presented
by Kniffen et al. (1995). The derived slope of the GRB spectrum above 30 MeV 
($\Gamma \simeq 2.1$) is consistent with that reported by Osborne et al. (1994),
while the normalization at 100 MeV is about a factor 1.3 higher.
The solid line is the sum of three distinct
 contributions (dashed lines): the fit to the X--ray background (XRB) obtained
with the AGN model of Comastri et al. (1995) and the FSRQ contributions to 
the GRB derived in the present model. 
It should be immediately noted that the fraction
of FSRQ with an MeV excess cannot account for more than about 10--20\%
 of the
``MeV bump" intensity, unless their number is largely underestimated and/or
their evolution properties are different. {\it It should be remarked, however,
 that
the new results presented by the COMPTEL collaboration at the conference in
 which this work has been submitted appear to convincingly disprove the 
existence of the ``MeV bump". In Fig.1 we have plotted the COMPTEL results
from the paper of Kappadath et al. (1995). It is seen that they are in very
good agreement with the intensity derived in our model}.

The contribution of FSRQ to the GRB is of the order of 70--80 \% 
that derived from the EGRET data at 100 MeV. 
With the
model's parameters it corresponds to a local emissivity at 100 MeV of
$3.8\times 10^{16}$ W Hz$^{-1}$ Gpc$^{-3}$, that is to the assumption of
an average broad spectral index $<\alpha_{x\gamma}>$ = 0.66 which is a
 reasonable compromise between the observed mean value of EGRET identified
FSRQ, $<\alpha_{x\gamma}> \simeq 0.57$, 
and the observational bias towards finding
objects with flatter $\alpha_{x\gamma}$.

It should be noted that the contribution of FSRQ to the XRB becomes appreciable
only above a few hundred keV and, therefore, it does not affect the XRB fit
obtained with the AGN model of Comastri et al. (1995).
 
 The FSRQ intensity resulting at 1 keV with our model is consistent with
the extrapolation, at the same energy, of the total flux due to the 
flat spectrum
radio sources at 5 GHz (i.e. 360 Jy sr$^{-1}$)
calculated by Setti \& Woltjer (1994). 
This extrapolation has been made with a computed $<\alpha_{rx}>$ = 0.91
and 
considering that the BL Lac contribution to the total flux is no more
than 5\%, as can be calculated using the radio fluxes at 5 GHz given
by Dondi \& Ghisellini (1995) and remembering that BL Lacs do not show any
evidence of cosmological evolution.
Another 10\% contribution is likely to be due to different
type of sources.
 
\par
In order to make full use of the available data we have made an
attempt to predict a Log N--Log S relationship and a redshift distribution and
compare them with those obtained with EGRET.
For this purpose we actually need a $\gamma$--ray luminosity function. Since
the published information does not allow us to compute it on the basis of
$\gamma$--ray data only, we have followed the approach of Padovani et al.
 (1993)
based on the existence of a strict correlation between the radio and 
$\gamma$--ray luminosities of blazars.  
First we have investigated the correlation between the radio
and the $\gamma$--ray luminosities (K-corrected) of EGRET FSRQ
finding a best fit relation $L_{\gamma} \propto L_{R}^{1.36 \pm 0.33}$.
Then we have considered the FSRQ luminosity
function given by Maraschi \& Rovetti (1994) as:
$dN/dL_{r} \propto L_R^{-\gamma}$, with $\gamma \simeq 2.4$. It follows
that the derived $\gamma$--ray luminosity function is a power law
with a flatter slope $\gamma \simeq 2.0$.
However the predicted redshift distribution is not fully consistent with the
observed one.

\par
A better approximation to the observed redshift distribution (Figure 2) has been
 obtained with a slightly different luminosity
function: a broken power law  with a slope of 1.5 above a minimum
luminosity of $5 \times 10^{45}$ ergs s$^{-1}$
and before a break luminosity
of $3 \times 10^{47}$ ergs s$^{-1}$ and a slope 2.4 up to a maximum luminosity
of $10^{49}$ ergs s$^{-1}$, while the normalization is obtained by the
 consistency with the local emissivity at 100 MeV as given above.
We note that some indication of a flattening
towards low luminosities is also present in the
Maraschi \& Rovetti (1994) radio luminosity function.
The expected source counts are presented in Figure 3, together with the
observational data where we have taken the maximum observed fluxes as reported
in Thompson et al (1995). All quoted values concern 10.4 sr outside the galactic
plane ($b > \pm 10^o$) as in Chiang et al. (1995). A straightforward comparison
of the predicted counts with the observed ones is obviously difficult given
the source time variability.

\par
We note that if the observed point at the brightest end of the Log N -- Log S
(13 objects all togheter) would be moved to a lower flux limit
by an average factor of two, to somehow correct for the time variability 
of the sources, it would still be consistent within 1$\sigma$ with
our predicted source counts. On the other hand 
it is clear that the sky--coverage at fainter fluxes must be progressively 
much lower, 
as
suggested by the flattening of the observed counts.
In fact a $V/V_{max}$ test assuming an homogeneous sky--coverage
at the faintest detected flux provide a mean value of
$\simeq 0.33$, suggesting either a negative evolution
(which can be excluded on the basis of radio and X--ray 
evolutionary properties) or a loss of sources at faint 
EGRET fluxes.
This is also consistent with the redshift distribution (Figure 2)
where it is seen that the observed number of objects in the
first redshift bin is much lower than the predicted one: low luminosity
fainter objects are mostly contributing to this redshift bin.

%
   \begin{figure}[htbp]
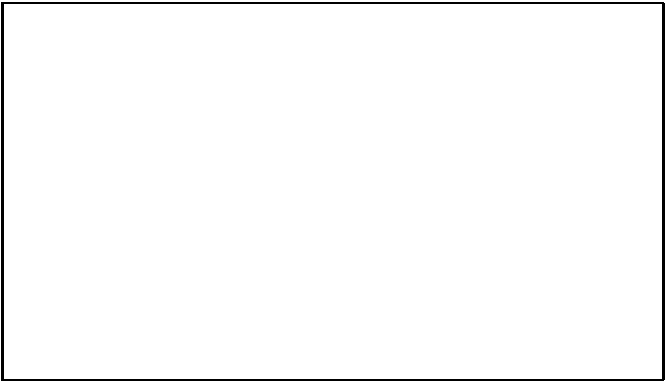

      \picplace{5cm}
      \caption{ The observed redshift distribution (solid line) of the 
               so far identified EGRET FSRQ (Thompson et al. 1995) 
               compared with the predicted distribution (dashed line).
                   }
         \label{FigVibStab}
   \end{figure}
%
%

%
   \begin{figure}[htbp]
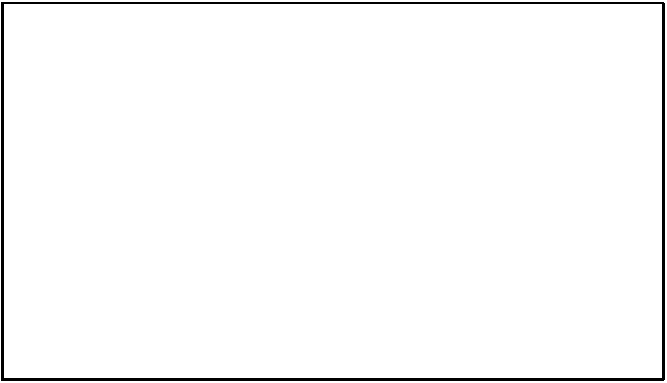

      \picplace{5cm}
      \caption{The predicted Log N -- Log S distribution for FSRQ
               compared with EGRET observations, adopting the 
              maximum fluxes reported by Thompson et al. 1995.
                            }
         \label{FigVibStab}
   \end{figure}
%
%

\section{Conclusions}

The $\gamma$--ray properties of FSRQ are such that they can account
for most of the GRB above 30 MeV but not for the ``MeV bump".
To this end one would require a large fraction of the FSRQ to be
``MeV blazars" which is contrary to the findings obtained
by the combined COMPTEL and EGRET observations.
The predicted MeV background intensity is a factor
from 5 to 10 below the fit of the ``MeV bump" given by
Gruber (1992) and it may very well account for the much lower
background intensity presented by the COMPTEL collaboration.

\par
From the radio luminosity function of the FSRQ and its time evolution
we have derived a $\gamma$--ray luminosity function
at 100 MeV which is consistent with present constraints on the
total number of FSRQ and their redshift distribution obtained
with the EGRET survey of the $\gamma$--ray sky.
However we note that the unknown EGRET survey sky--coverage
  prevents us from a detailed comparison between the observational 
data and model predictions.
\par
As a general conclusion it appears that various classes of AGN are
able to account for most of the extragalactic background radiations
observed over the very wide energy interval from about 1 keV to tens
of GeV.

\begin{acknowledgements}
    This work has been partially supported by the Italian
     Space Agency (ASI--94--RS--96)
\end{acknowledgements}

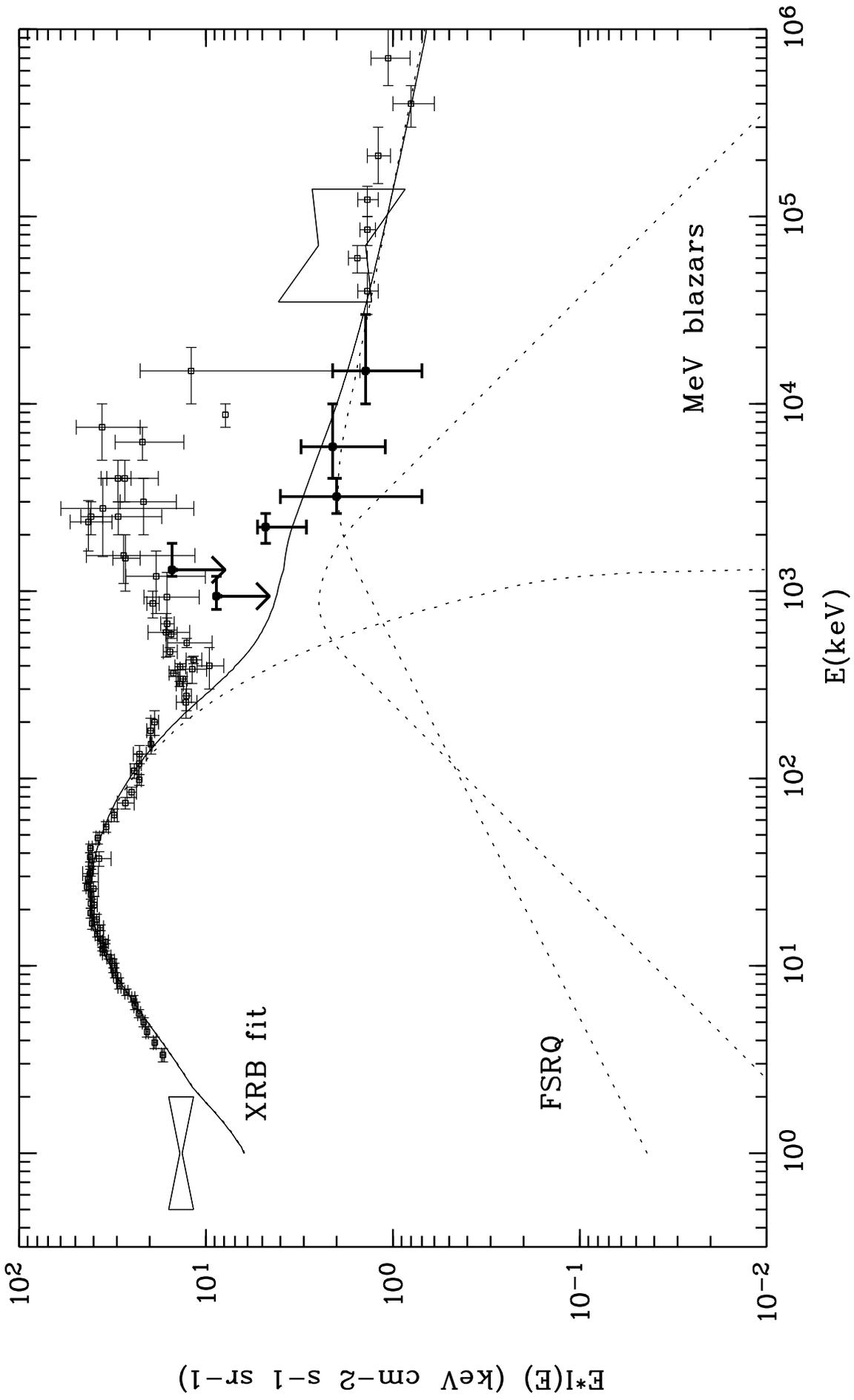
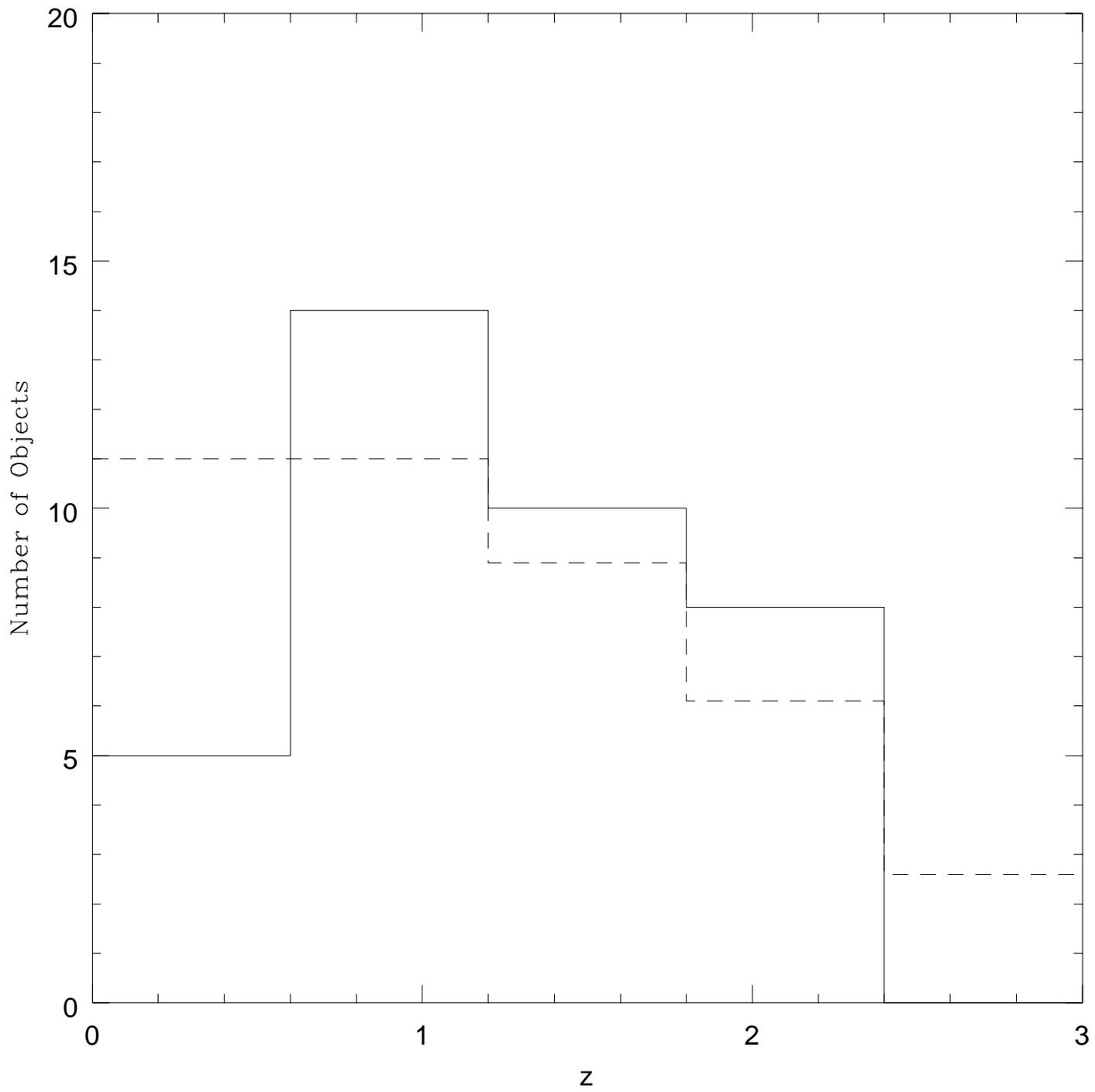
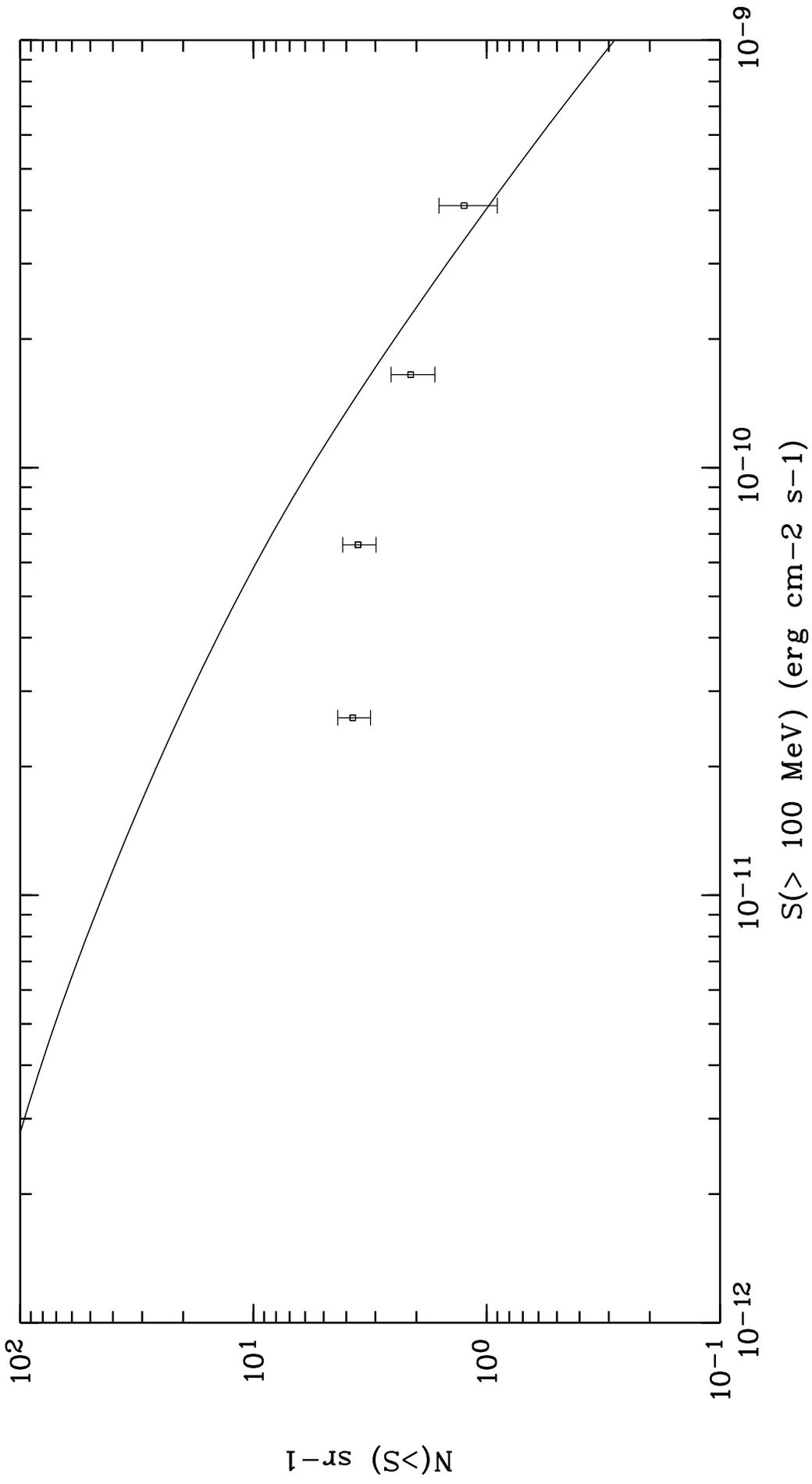
\end{document}